\documentclass[twocolumn,pre,floats,floatfix,aps,amssymb]{revtex4}
\usepackage{amsmath}
\usepackage{nicefrac}
\usepackage{amssymb}
\usepackage{graphicx}
\usepackage{amsfonts}
\usepackage{color}
\usepackage{epstopdf}
\DeclareGraphicsExtensions{.pdf,.eps,.png,.jpg,.mps}
\begin{document}

\title{The emergence and analysis of Kuramoto-Sakaguchi-like models as an effective description for the dynamics of coupled Wien-bridge oscillators}

\author{L.Q. English$^1$, David Mertens$^2$, Saidou Abdoulkary$^{1,3,4}$, C.B. Fritz$^1$, K. Skowronski$^1$, P.G. Kevrekidis$^5$}

\affiliation{$^1$ Department of Physics and Astronomy, Dickinson College, Carlisle, Pennsylvania, 17013, USA \\
$^2$ Department of Physics, Eckerd College, St. Petersburg, FL 33711, USA \\
$^3$ D\'epartement des Sciences Fondamentales, de Droit et des Humanit\'es, IMIP University of Maroua, P.O. Box 46, Maroua, Cameroon \\
$^4$ Laboratory of Mechanics, Department of Physics, University of Yaound\'e I, Cameroon \\
$^5$ Department of Mathematics and Statistics, University of Massachusetts, Amherst, MA 01003}

\begin{abstract}
We derive the Kuramoto-Sakaguchi model from the basic circuit equations governing two coupled Wien-bridge oscillators. A Wien-bridge oscillator is a particular realization of a tunable autonomous oscillator that makes use of frequency filtering (via a RC band-pass filter) and positive feedback (via an Op-Amp). In the last few years, such oscillators have started to be utilized in synchronization studies. We first show that the Wien-bridge circuit equations can be cast in the form of a coupled pair of Duffing - Van der Pol equations. Subsequently, by applying the method of multiple time scales, we derive the differential equations that govern the slow evolution of the oscillator phases and amplitudes. These equations are directly reminiscent of the Kuramoto-Sakaguchi type models 
for the study of synchronization.  We analyze the resulting 
system in terms of existence and stability of various coupled oscillator solutions and explain on that basis how their synchronization 
emerges. The phase-amplitude equations are also compared numerically to the original circuit equations, and good agreement is found. Finally, we report on  experimental measurements on two coupled Wien-bridge oscillators and relate the results back to the theoretical predictions. 
\end{abstract}

\maketitle

\section{Introduction}
The Kuramoto model was originally introduced by Yoshiki Kuramoto as a mathematically tractable way of explaining the synchronization observed in biological systems \cite{kura, strogatz}. The model and its various modifications, however, are generic, in the sense that they do not start with biological or physical 
first principles. Instead, they represent a type of ``normal form'', a
 prototypical model system that captures the mathematical essence of the
phenomenon.
Consequently, they do not aim to directly describe the details of any real physical or biological systems or processes. Nonetheless, these phase models have had remarkable success as effective models of synchronization phenomena in 
a diverse range of contexts~\cite{acebron}. 

The literature is replete with theoretical and numerical predictions for synchronizing nonlinear oscillators (see, for instance, the book of~\cite{Pikov}
for a relevant summary). These predictions are almost always written in non-dimensionalized forms, but for many experimental systems, equations from first principles do not easily transform into such canonical non-dimensionalized equations. Establishing such a connection, however, between the experimental parameters and the coefficients in the canonical equations, can be quite powerful, as it provides a wealth of predictive power for the particular experimental system,
given the breadth and depth of associated theoretical analysis and also 
the amenability of such systems to computational studies. 
Here, we will attempt to establish such a connection between experiments,
numerical computations (both direct dynamics and bifurcation-type
analysis) and analytical theory (based on multiple time scales analysis)
for a nonlinear electrical circuit system.

In this paper, more specifically, we derive the connection from component-level system specifications to the canonical phase and amplitude equations for Wien-bridge electronic self-oscillators.  In a small number of other systems, such as coupled Josephson arrays \cite{wiesenfeld} and mechanical oscillators \cite{mertens, pantaleone}, it has been possible to mathematically derive the effective phase model starting from basic principles. Such scientifically relevant systems, however, seem to be quite rare in the literature. In these cases, a Kuramoto-like model emerges after a procedure that suitably averages out the fast time dynamics. In this paper, too, we use a similar idea to bridge the two seemingly disparate levels of description - the description at the level of the Op-amps in terms of voltages and currents, and the higher-level description in terms of oscillator phases and amplitudes. 

We analyze the behavior of two coupled weakly nonlinear oscillators. The slow dynamics obtained using the method of multiple-time scales can be separated into amplitude and phase dynamics. When the amplitudes are constant, the system reduces to a case example of the phase dynamics  within the 
well-studied Kuramoto-Sakaguchi model \cite{saka}, and we can make specific 
predictions for parameters appearing in that model in terms of circuit 
component values. We also investigate the dynamics when the amplitudes are 
not assumed constant and obtain systems similar in form to those rigorously studied by Aronson {\it et al.}~\cite{aronson}. We further generalize the equations by examining the case where the two oscillators are comprised of circuit element of non-identical values. Finally, in all of these cases, 
we compare the theoretical predictions to experimental measurements on a pair of coupled Wien-bridge oscillators. We find good agreement in the parametric
regimes where the reduction is expected to be valid (and discuss the relevant
deviations when it is not).

A previous study~\cite{english} found experimentally that the dynamics of high-gain Wien-bridge oscillators was described quite well by the Kuramoto-Sakaguchi phase model. In this paper we focus on weakly nonlinear oscillators, in contrast to strongly nonlinear oscillators studied there. We also explore 
the synchronization dynamics from a bifurcation theory perspective
and identify the potential of the model for spontaneous symmetry breaking
features (even though we identify the latter as unstable).  
Although the weakly nonlinear oscillators are nearly simple 
harmonic oscillators, their coupled behavior is not well described as a pair of harmonic oscillators; the latter description is inadequate to 
characterize the dynamical features we observe. Though small, 
the nonlinear terms stabilizing the oscillator amplitudes lead to behavior on two different time scales, the fast oscillation rate and the slow coupled amplitude dynamics. The method of multiple-time scales systematically 
decouples the analysis of the two time scales, and provides the crucial 
connection between the component-level description of the oscillators and the 
effective phase/amplitude reduction model. The latter provides the path
to the synchronization and spontaneous symmetry breaking features observed
and paves the way towards the analysis at the lattice level of a large
number of associated oscillators. For completeness, but also
given the interest of the latter theme in its own right, we also
consider the more technically challenging (as regards the
derivation of the slow dynamics) scenario of two 
non-identical oscillators.

Our presentation will be structured as follows. In section II, we 
will start from the first principles of the electrical circuit system
of interest and will derive the amplitude/phase effective model via
multiple scales analysis. In section III, we will obtain theoretical
and numerical conclusions on the basis of the latter model, which
then in section IV will be compared to the direct physical experiments.
Finally, section V summarizes our findings and presents our conclusions,
as well as some directions for potential future work. Technical details of the 
multiple scale analysis can be found in the Appendix.

\section{Modeling Analysis: From Electrical Circuits to Amplitude and Phase
Descriptions}

%In this section we derive amplitude oscillator equations for our interacting electronic oscillators. 
Our first-principles derivation will begin with basic circuit equations under the assumption of ideal circuit elements. The mid-point of our calculation will be a pair of coupled second-order differential equations describing the dynamics of the Op-Amps' input voltages in terms of component values and input voltages. 
From there, the multiple-scale analysis will lead to 
four coupled first-order differential equations describing the dynamics in terms of amplitude and phase of the pair of coupled oscillators under consideration.

\subsection{Two identical coupled Wien-bridge oscillators and their circuit equations}

\begin{figure}
\begin{center}
\includegraphics[width=3.2in]{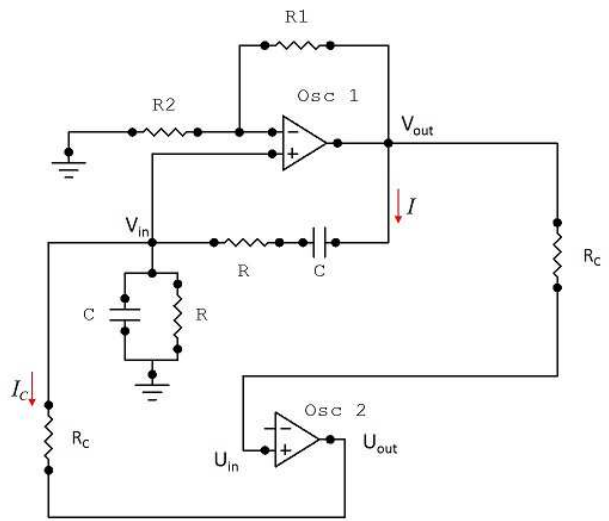}
\caption{The circuit diagram of the coupled Wien-bridge oscillators. Note that the second oscillator, labeled Osc2, is not shown in its full detail. The resistors labeled $R_c$ couple the two oscillators to one another.}
\label{diagram}
\end{center}
\end{figure}

The basic circuit diagram is shown in Fig.~\ref{diagram}. The top oscillator is shown with all of its necessary components. We see the non-inverting amplifier loop involving the negative terminal of the Op-Amp. Note that in an actual Wien-bridge oscillator we also need some nonlinear circuit element to stabilize the oscillations. One easy way to incorporate this aspect is to make $R_1$ slightly voltage-dependent, such that for large voltages, $R_1$ would be reduced. In previous experiments, a diode pair in parallel with $R_1$ is used to accomplish this \cite{english}, and we will take that approach in the analyses and experiments presented here. The circuit diagram also features positive feedback (from the Op-Amp output to the positive input channel) via an RC bandpass filter. This filter consists of a resistor, $R$, and capacitor in series, followed by a resistor and capacitor in parallel to ground. 

We have highlighted in the figure some notation that we will use in the derivation of the circuit equations. For instance, the voltages are labeled at particularly important junctions, such as the output terminals of the Op-Amps, denoted by $V_{out}$ and $U_{out}$ for oscillators 1 and 2, respectively. Also note that the variables with subscript `c' refer to the coupling between the two oscillators; $R_c$ is the coupling resistor, and $I_c$ the coupling current.

We will begin by considering the Kirchhoff junction rule at the label $V_{in}$ in Fig.~\ref{diagram}. The currents flowing to ground through the parallel resistor and capacitor are $V_{in}/R$ and $C \dot{V}_{in}$, respectively, while the current flowing from $V_{in}$ to $U_{out}$ is $I_c = \left(V_{in} - U_{out}\right) / R_c$. As such we have
\begin{equation}
I=\frac{V_{in} - U_{out}}{R_c} + C \dot{V}_{in} + \frac{V_{in}}{R}.\label{eq:V_in_junction}
\end{equation}
The voltage drop from the first oscillator's output to input terminals is $V_{cap} + I\,R$ (where the subscript stands for capacitor). Using Eq.~(\ref{eq:V_in_junction}), this becomes
\begin{equation}
V_{out} - V_{in} = V_{cap} + \left(\frac{V_{in} - U_{out}}{R_c} + C \dot{V}_{in} + \frac{V_{in}}{R}\right)R
\label{eq:V_out_V_in_no_I}
\end{equation}
Although we do not know the voltage across the capacitor, we know that the current can be related to the derivative of the capacitor's voltage as $I=C\dot{V}_{cap}$. Taking the derivative of Eq.~(\ref{eq:V_out_V_in_no_I}) and eliminating $\dot{V}_{cap}$ leads to
\begin{align}
0= & R\,C\frac{d^{2}V_{in}}{dt^{2}}+3\frac{dV_{in}}{dt}+\frac{V_{in}}{R\,C}+\frac{1}{R_{c}C}\left(V_{in}-U_{out}\right) \nonumber \\
& + \frac{R}{R_{c}}\frac{d}{dt}\left(V_{in}-U_{out}\right)-\frac{dV_{out}}{dt}.
\label{eq:almost_main}
\end{align}
This describes the coupled dynamics of our oscillators using two voltages for each oscillator: the Op-Amp output and the non-inverting input. We will next eliminate one of these.

The output voltage in Eq.~(\ref{eq:almost_main}) needs to be replaced with an expression involving the input voltage. To obtain this, consider the non-inverting amplifier block of oscillator 1, the portion involving the resistors $R_{1}$ and $R_{2}$. For ideal Op-Amps, the current entering the inverting input of the 
Op-Amp is negligible. Furthermore, the Op-Amp is in a negative feedback configuration, ensuring that the non-inverting and inverting inputs will be maintained at essentially the same voltage, $V_{in}$. As such, $V_{out}$ and $V_{in}$ are related via,
\begin{equation}
V_{out}=\left(1+\frac{R_{1}}{R_{2}}\right)V_{in}
\end{equation}
Recall that $R_1$ will need to be nonlinear for stable oscillations, and that the circuit involves diodes wired in parallel with $R_1$. For voltages near the diode threshold, the diodes allow current to bypass $R_{1}$; equivalently, the resistance of $R_{1}$ drops for large voltages. To first order in small quantities\footnote{The squared voltage in equation \ref{nonlin} should be the difference $V_{out}-V_{in}$, not simply $V_{in}$. However, to first order in small quantities, this merely results in a redefinition of $\varepsilon_v$.}, we can approximate this effect as
\begin{equation}
R_{1}\simeq R_{01}\left(1-\varepsilon_{v}\,V^{2}_{in}\right).\label{nonlin}
\end{equation}
The resistance should be an even function of voltage (or current), and so Eq.(\ref{nonlin}) could also be thought of as the first two terms in a Taylor expansion. Similarly, it is clear that $\varepsilon_{v}$ is positive. The value of $\varepsilon_v$ could in principle be calculated using the Shockley diode equation \cite{shockley}, but for our purposes we will simply assume it is a small positive number. It then follows that 
\begin{equation}
V_{out}=\left(1+\frac{R_{01}}{R_{2}}\left(1-\varepsilon_{v}\,V_{in}^{2}\right)\right)V_{in},\label{approx}
\end{equation}
where $R_{01}$ is the resistence of the actual resistor component in the circuit. This approximation is valid in the weakly nonlinear regime, which means the current bypassing the resistor via the diodes must be small. The same equation will also govern $U_{out}$.

Substituting this result into Eq.~(\ref{eq:almost_main}) to eliminate
$V_{out}$ and $U_{out}$, we finally obtain the following second-order
differential equation for oscillator 1: 
\begin{align*}
R\,C\frac{d^{2}V_{in}}{dt^{2}}+\frac{dV_{in}}{dt}\left(2+\frac{R}{R_{c}}-\frac{R_{01}}{R_{2}}+3\varepsilon_{v}\frac{R_{01}}{R_{2}}V_{in}^{2}\right)+ \\
V_{in}\left(\frac{1}{R\,C}+\frac{1}{R_{c}C}\right)-\frac{U_{in}}{R_{c}C}\left(1+\frac{R_{01}}{R_{2}}-\varepsilon_{v}\frac{R_{01}}{R_{2}}U_{in}^{2}\right) \\
-\frac{dU_{in}}{dt}\left(\frac{R}{R_{c}}\right)\left(1+\frac{R_{01}}{R_{2}}-3\varepsilon_{v}\frac{R_{01}}{R_{2}}U_{in}^{2}\right)=0.
\end{align*}
This can be simplified by introducing a number of parameters denoting
dimensionless time and important ratios: 
\begin{align*}
\tau & \equiv\frac{t}{R\,C}, & \varepsilon_{c} & \equiv\frac{R}{R_{c}}, & \varepsilon_{g} & \equiv\frac{R_{01}}{R_{2}}-2.
\end{align*}
Here we have made the notational implicit assumption that $R_{c}\gg R$, i.e.
the coupling is weak. Our definition for $\varepsilon_{g}$ is motivated
by the requirement for stable oscillations (for a single oscillator), namely that the Op-Amp gain should be slightly larger than 3. This implies that $0 < \varepsilon_g << 1$. %with a small damping or driving. 
Using primes to denote differentiation with respect to $\tau$,
this becomes
\begin{align}
&V_{in}^{\prime\prime}+V_{in}^{\prime}\left(-\varepsilon_{g} +\varepsilon_{c}+3\,\left(2+\varepsilon_{g}\right)\,\varepsilon_{v}\,V_{in}^{2}\right)+  \nonumber \\
& V_{in}\left(1+\varepsilon_{c}\right)-U_{in}\varepsilon_{c}\left(3+\varepsilon_{g}-\left(2+\varepsilon_{g}\right)\,\varepsilon_{v}\,U_{in}^{2}\right) \nonumber \\
&-U_{in}^{\prime}\varepsilon_{c}\left(1+\left(2+\varepsilon_{g}\right)-3\,\left(2+\varepsilon_{g}\right)\,\varepsilon_{v}\,U_{in}^{2}\right)=0 \nonumber \\
\label{eq:governing-eqs}
\end{align}
When all $\varepsilon$ are sufficiently small, then the governing equation is well approximated by,
\begin{align}
&V_{in}^{\prime\prime}+V_{in}^{\prime}\left(-\varepsilon_{g} +\varepsilon_{c}+6\,\varepsilon_{v}\,V_{in}^{2}\right)+V_{in}\left(1+\varepsilon_{c}\right)- \nonumber \\
&3U_{in}\varepsilon_{c}-3U_{in}^{\prime}\varepsilon_{c} \approx 0.
\label{eq:governing-eqs2}
\end{align}
Here we have dropped all products of small quantities.
The limit $\varepsilon\to0$ corresponds to uncoupled oscillators
with linear $R_{1}$, in which case we have simple harmonic oscillators.
Since everything is symmetric between the two oscillators, the equation
for the oscillator 2 will be identical in form, with $U$ and $V$ interchanged.

If we let the coupling parameter, $\varepsilon_c$, go to zero in Eq.( \ref{eq:governing-eqs2}), we obtain a Duffing - Van der Pol oscillator. Note that the sign of the damping coefficient depends on the amplitude of the voltage: for low voltage values the coefficient is negative (amplification) but for larger voltages it becomes positive (dissipation).   

\subsection{Multiple Time Scales}
The system of equations given in Eq.~(\ref{eq:governing-eqs}) can be simulated directly, of course, but first we choose a different approach, since the larger goal is to establish a mathematical connection with the Kuramoto-Sakaguchi 
model and to use the latter for a series of quantitative conclusions about
our system. Such models describe the dynamics in terms of oscillator phases, as well as possibly oscillator amplitudes. In the original Kuramoto model, the phases evolve according to the natural frequencies of the oscillators which are modulated by phase interaction terms whose strength is given by the coupling parameter. Thus, the typical problem has two time scales: a fast dynamics governed by the natural oscillator frequencies, and a slower dynamics representing the phase interactions between the coupled oscillators. This suggests that the method of multiple time-scales is a fruitful approach towards this problem \cite{Nayfeh, strogatz2}.

We now sketch the analysis using the two-timing method, deferring technical
details to the appendix. The starting point is the following expansion of the voltages appearing in our governing
circuit equations, namely Eq.~(\ref{eq:governing-eqs}),  
\begin{align}
V_{in} & =V_{0}(T_{0},\,T_{1})+\varepsilon_{1}\,V_{1}(T_{0},\,T_{1})+\dots\nonumber \\
U_{in} & =U_{0}(T_{0},\,T_{1})+\varepsilon_{1}\,U_{1}(T_{0},\,T_{1})+\dots\label{eq:ansatz}
\end{align}
where $T_{0}\equiv\tau$ and $T_{1}\equiv\varepsilon_{t}\,\tau$ are
the fast and slow time-scales, respectively. The perturbation analysis suggests the following form (see appendix):
\begin{align}
V_0(T_0,T_1)=A_v(T_1)\, e^{\imath T_0} + C.C.,
\label{ansatz}
\end{align}
where C.C. stands for complex conjugate.
We are interested in the time evolution of the slow-varying complex amplitudes, $A_v \equiv \frac{1}{2}a_v e^{\imath \phi_v}$ and $A_u \equiv \frac{1}{2}a_u e^{\imath \phi_u}$. Following the two-time-scale procedure detailed in the appendix, we finally arrive at the coupled phase-amplitude evolution equations:
\begin{align}
\dot{a}_{v} =&\frac{3\sqrt{2}\varepsilon_{c}}{2} a_{u}\,\cos\left(\phi_{u}-\phi_{v}-\frac{\pi}{4}\right) \nonumber \\
&-\frac{2\varepsilon_{c}-2\varepsilon_{g}+3\varepsilon_{v}a_{v}^{2}}{4} a_{v}, \nonumber \\
a_{v}\,\dot{\phi}_{v} =&\frac{3\sqrt{2}\varepsilon_c}{2} a_{u}\,\sin\left(\phi_{u}-\phi_{v}-\frac{\pi}{4}\right)+\frac{\varepsilon_{c}}{2} a_{v}.
\label{result_b}
\end{align}
The equations for the other oscillator with variables $a_u$ and $\phi_u$ are obtained by simply swapping the subscripts $u$ and $v$ in Eq.~(\ref{result_b}).

If the oscillator amplitudes approach a steady-state
value, and no symmetry-breaking transition occurs, then we can assume $a_{v}=a_{u}$, and the dynamics for the phases is given by,
\begin{align}
\dot{\phi}_{v}&=\frac{1}{2}\varepsilon_{c}+\frac{3\sqrt{2}}{2}\varepsilon_{c}\sin\left(\phi_{u}-\phi_{v}-\frac{\pi}{4}\right) \nonumber \\
\dot{\phi}_{u}&=\frac{1}{2}\varepsilon_{c}+\frac{3\sqrt{2}}{2}\varepsilon_{c}\sin\left(\phi_{v}-\phi_{u}-\frac{\pi}{4}\right).
\label{SK_final_b}
\end{align}
This result has the Sakaguchi-Kuramoto form~\cite{saka} 
for a specific choice of their parameter $\alpha=\frac{\pi}{4}.$

\begin{figure}
\begin{center}
\includegraphics[width=3.4in]{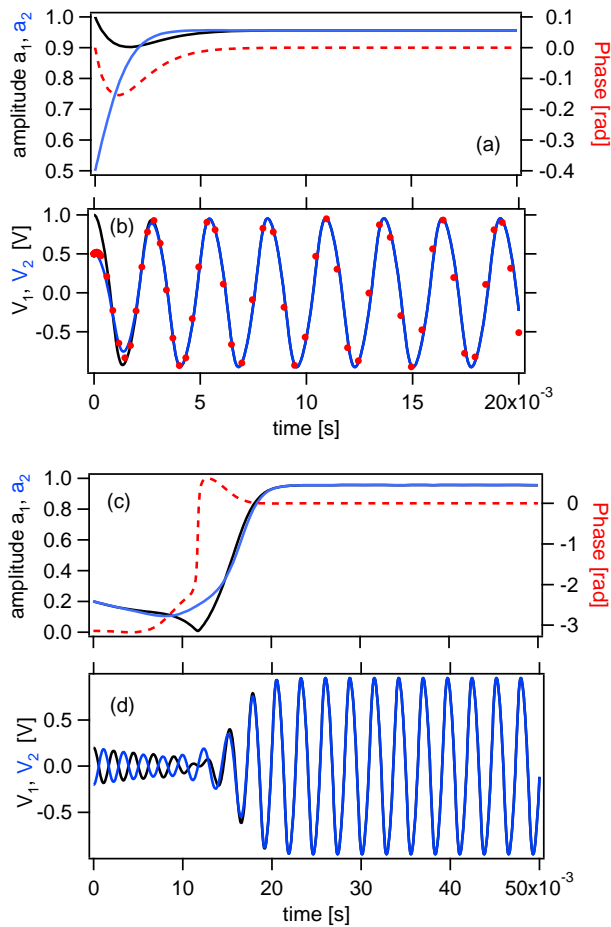}
\caption{Numerical comparison of the derived amplitude-phase equations
of  Eq.~(\ref{result_b}) to the governing circuit equations of Eq.~(\ref{eq:governing-eqs}). The parameters here are: $\varepsilon_c=0.065$, $\varepsilon_v=0.24$, and $\varepsilon_g=0.22$. (a) and (b): initial conditions of $\theta=0$, $a_1=1$, and $a_2=0.5$. The dashed (red) trace in (a) corresponds to the phase difference (right axis). The (red) markers in (b) is the voltage of the second oscillator computed form the phase-amplitude predictions in (a). (c) and (d): initial conditions of $\theta=\pi$.}
\label{sims}
\end{center}
\end{figure}

To qualitatively assess the accuracy of the amplitude-phase equations, we compared simulations of Eq.~(\ref{result_b}) with those of the electrical oscillator dynamics, Eq.~(\ref{eq:governing-eqs}). The results are shown in Fig.~\ref{sims}. The upper two panels correspond to initial conditions where the voltages start in phase with each other but at different amplitudes, and the lower two panels are for equal amplitude but anti-phase initial conditions. Furthermore, panels (a) and (c) display the simulated  amplitudes (left axis) and phase difference (right axis) using Eq.(\ref{result_b}), whereas panels (b) and (d) show the actual voltage oscillations computed from Eq.~(\ref{eq:governing-eqs}). The (red) dots in Fig. \ref{sims} (b) depict the voltage time series of $V_2$ reconstructed from the computed amplitude and phase shown in (a). A similar reconstruction was done for the initial conditions in (c) and (d), but are not shown. The simulations compare favorably.

The amplitude-phase equations give excellent steady-state results. The amplitudes and oscillation rates closely match those of the voltage simulations, even though the initial conditions vary significantly. The primary failure of the amplitude-phase equations occurs when the amplitudes become very small. As seen in panel (c), the small amplitudes lead to excessively large phase velocities. These excessive phase velocities are not a numerical artifact as they persist even for very small integrator time steps. As such, we will limit our analysis to steady-state results, where the behavior is recovered almost perfectly by the amplitude-phase equations.

\section{Theoretical and Numerical Results}

We now turn our attention to extracting predictions that can be directly compared against our experiments.

The two-oscillator problem is well studied, the most complete analysis being that of Aronson and coworkers \cite{aronson}. Aronson's paper presents a detailed analysis of a pair of oscillators near a Hopf bifurcation. Unfortunately, they focus on two sets of simplifications that limit how we can apply their findings. First, the bulk of their analysis is for ``scalar'' coupling, which is inconsistent with the negative $\nicefrac{\pi}{4}$ phase offset in Eq.~\ref{result_b}. Second, the small portion of their paper that does not assume scalar coupling focuses predominantly on coupling with ``sheer.'' Our system has zero sheer. Apparently, our oscillators' specific set of conditions have not been carefully analyzed before.

\subsection{Primary steady-state behavior}
We will begin by considering oscillators with identical and nonzero amplitudes, i.e. $a_u = a_v$. The phase dynamics then become independent of the amplitude dynamics and we can think of our system as merely a pair of phase oscillators. We have already obtained the phase dynamics in Eq.~(\ref{SK_final_b}).

The phase dynamics of Eq.~(\ref{SK_final_b}) contain two fixed points for the relative phase. To see this, we write
\begin{align}
\frac{d}{dt}\left(\phi_v-\phi_u\right)=&{}\frac{3\sqrt{2}}{2}\varepsilon_c \left[\sin\left(\phi_u-\phi_v-\pi/4\right)\right. \nonumber \\ 
&-\left.\sin\left(\phi_v-\phi_u-\pi/4\right)\right], 
\label{phasediff}
\end{align}
Define $\theta \equiv \phi_v-\phi_u$, a quantity representing the phase difference between the two oscillators. We can then write Eq.~(\ref{phasediff}) more concisely as
\begin{equation}
\dot{\theta} = -3\varepsilon_c \sin \theta.
\label{theta1}
\end{equation}
The relative phase clearly has a stable fixed point of $\theta=0$ and an unstable one of $\theta=\pi$. When the amplitudes are identical, we expect a steady-state in which the two oscillators have the exact same phase.

It is noteworthy that Eq.~(\ref{theta1}) can be integrated to yield an exact solution,
\begin{equation}
\theta (t) = 2\tan ^{-1}\left(\tan \frac{\theta_0}{2} e^{-3\varepsilon_c t}\right),
\label{theta_exp}
\end{equation}
where $\theta_0$ is the initial phase difference at $t=0$. Unfortunately this prediction is of little practical use. As we show in the coming discussion, if two oscillators are prepared with a nonzero phase difference, their amplitudes will not remain identical even if they evolve toward a steady state with identical amplitudes.

In the expected steady-state, the synchronization frequency depends on the strength of the coupling. Inserting $\theta=0$ into Eq.~(\ref{SK_final_b}) leads to
\begin{align}
\dot{\phi}=\frac{\varepsilon_c}{2}+\frac{3\sqrt{2}}{2}\varepsilon_c \sin\left(-\frac{\pi}{4}\right) = - \varepsilon_c.
\label{predic}
\end{align}
Recall that $\phi$ represents deviations from the intrinsic, fast oscillations. We would therefore expect that increasing $\varepsilon_c$ would decrease the synchronization frequency.

Consider the amplitude dynamics within Eqs.~(\ref{result_b}). In the absence of coupling ($\varepsilon_c=0)$, the steady-state amplitude is given by $a_0=\sqrt{\nicefrac{2\varepsilon_g}{3\varepsilon_v}}$. For the coupled case, we can again start from Eq.~(\ref{result_b}) together with $a_u=a_v$. Employing a trigonometric identity, we obtain the following differential equation governing the amplitude:
\begin{align}
\dot{a}_v=\frac{3}{2}\varepsilon_c\left(\cos\theta+\sin\theta\right)a_v-\frac{1}{2}\left(\varepsilon_c-\varepsilon_g\right)a_v-\frac{3}{4}\varepsilon_v a_v^3.
\end{align}
At the expected steady-state, where $\theta=0$, the amplitude dynamics have the stable fixed point
\begin{align}
a_v^2 = \frac{4\varepsilon_c+2\varepsilon_g}{3 \varepsilon_v}.
\label{amp_pred}
\end{align}
If two coupled oscillator synchronize with equal amplitudes, then their coupled amplitudes will be larger than their uncoupled ones.

There is an additional noteworthy feature about the coupled amplitudes: with coupling in place, oscillations can be stable even for negative $\varepsilon_g$. Recall that individual uncoupled oscillators require $\varepsilon_g > 0$ in order for oscillations to start. For coupled oscillators, $\varepsilon_g$ can be as negative as $-2 \varepsilon_c$. For the interval $-2 \varepsilon_c < \varepsilon_g < 0$, we should observe oscillations when the two oscillators are connected via the coupling resistor, $R_c$, and those oscillations will cease when that resistor is removed. The coupling itself makes such oscillations possible, an instance of what we might call symbiotic oscillations also seen in chemical oscillations \cite{smale}. 

If we observe two identical oscillators with the same amplitudes, we expect that (a) the oscillators will have identical phases, (b) their coupled phase velocities will be slower than their uncoupled phase velocities, (c) their coupled amplitudes will be larger than their uncoupled amplitudes, and (d) there should be a regime in which the individual oscillators do not oscillate but the coupled pair does.

\subsection{Alternative states and stability analysis}
Equal-amplituded oscillations are the simplest and most obvious choice to consider, but other steady states may arise. In order to identify and assess the stability of other steady states, we now consider the possibility of distinct amplitudes.

If we treat the amplitudes more generally as independent dynamical variables, Eqs.~(\ref{result_b}) can be recast in the following compact form:

\begin{align}
{a_1}^\prime &= \frac{3}{2}(\cos \theta - \sin \theta) a_2 - \frac{1}{2}(1-r)a_1-\frac{3}{4}s a_1^3
\nonumber \\
{a_2}^\prime &= \frac{3}{2}(\cos \theta + \sin \theta) a_1 - \frac{1}{2}(1-r)a_2-\frac{3}{4}s a_2^3
\nonumber \\
\theta^ \prime &= -\frac{3}{2}\sin \theta\left(\frac{a_2}{a_1}+\frac{a_1}{a_2}\right)-\frac{3}{2}\cos \theta \left(\frac{a_2}{a_1}-\frac{a_1}{a_2}\right), 
\label{system}
\end{align}
where $s=\frac{\varepsilon_v}{\varepsilon_c}$ and $r=\frac{\varepsilon_g}{\varepsilon_c}$. Both $\omega_0$ and $\varepsilon_c$ were absorbed into a dimensionless time variable, and primes indicates differentiation with respect to this variable. Here we also have opted for a slight notational change: instead of using the subscripts $u$ and $v$ to denote the first and second oscillator, respectively, we now use the subscripts 1 and 2.

Let us look at the state where both coupled oscillators are quiescent, corresponding to $a_1=a_2\rightarrow 0$. The phase behavior is irrelevant, so we make the convenient choice $\theta = 0$. This is clearly a fixed point of Eq.~(\ref{system}). To ascertain the stability of this fixed point, examine the Jacobian of the reduced system of the two amplitude equations obtained by setting $\theta=0$:
\[
J=  \begin{bmatrix}
   \frac{1}{2}(r-1) & \frac{3}{2} \\ \frac{3}{2} & \frac{1}{2}(r-1)\       \end{bmatrix}
\]
For stability, the eigenvalues of this Jacobian must be negative. Equivalently, the trace must be negative and the determinant positive, which yields the condition that $r<-2$. The quiescent state becomes unstable when the gain setting exceeds $r=-2$, corroborating what we found in the previous section.

Next, let us investigate the stability of the stable state identified in the previous section: symmetric oscillations of $a_1=a_2=a$ and $\theta=0$. In order for this to be a fixed point, it is clear that $a^2=\frac{4}{3}(\frac{r}{2}+1)\frac{1}{s}$. (Real solutions require $r>-2$, so the stable quiescent state gives way to this synchronized state without overlap or bistability.) To determine the stability of this fixed point, we evaluate the Jacobian of the full system at this fixed point:
\[
J = \begin{bmatrix} -r-\frac{7}{2} & \frac{3}{2} & -\frac{3}{2}a \\
\frac{3}{2} & -r-\frac{7}{2} & \frac{3}{2}a \\
\frac{3}{a} & -\frac{3}{a} & -3 \end{bmatrix}
\]
This Jacobian has three eigenvalues:
($-(r+2)$, $-(8+r \pm \sqrt{(r+8) (r-4)})/2$). Requiring that these be negative yields as the condition for stability $r>-2$ (from the first eigenvalue). At this value
of the parameter, the amplitude evaluates to $a=0$ as indicated above, 
so the system exhibits a pitchfork bifurcation away from the quiescent 
state at the critical gain value of $r^*=-2$.

\begin{figure}
\begin{center}
\includegraphics[width=3.2in]{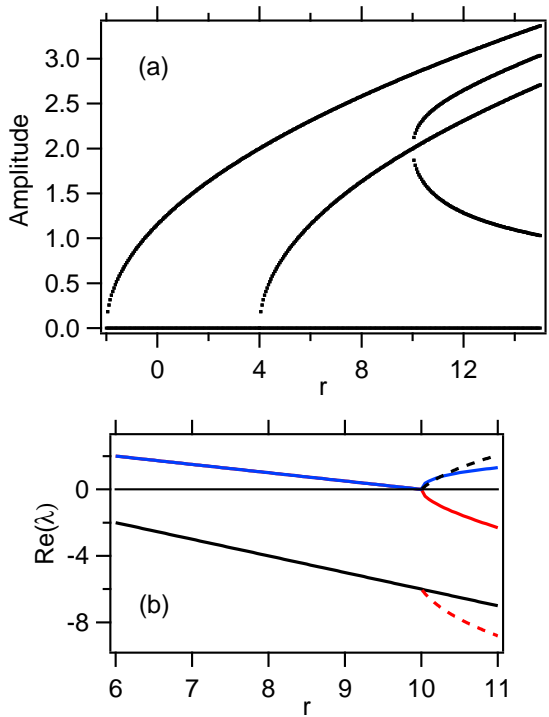}
\caption{(a) Numerical evaluation of fixed points of Eq.(\ref{system}) as a function of $r$ for fixed $s=1$. Note the unequal-amplitude curve that branches of the splayed state at $r=10$. (b) The Jacobian eigenvalues (real parts) associated with the splayed (solid, black) and unequal-amplitude (dashed, red) states.}
\label{gensol}
\end{center}
\end{figure}

A similar analysis is performed for the anti-symmetric (or splayed) fixed point, $a_1=a_2=b$ and $\theta=\pi$. Here, $b^2=\frac{4}{3}(\frac{r}{2}-2)\frac{1}{s}$, which means that the fixed point only emerges when $r>4$. 
We identify a complex conjugate pair of the eigenvalues of the linearization
around the splayed state as being responsible for its instability
for $4 < r < 10$. Beyond $r=10$, a remarkable and non-generic
bifurcation scenario arises. While the third eigenvalue remains
negative and real (as is shown in panel (b) of Fig.\ref{gensol}),
the complex conjugate pair approaches the origin and collides---as a 
pair---with the origin, giving rise to a pair of real eigenvalues. This is,
to the best of our knowledge, a non-generic example of a sub-critical
pitchfork bifurcation, as the symmetric state {\it remains} unstable
past this threshold of $r=10$, yet now it is because of a single positive 
real eigenvalue (instead of two complex eigenvalues with positive real
part for $r<10$), while the second eigenvalue splits along the negative
real axis. The pitchfork nature of this bifurcation, nonetheless,
suggests the existence of an additional (broken-symmetry) branch
which we now explore.

An intriguing feature of our system is that, in a way reminiscent of 
couplers extensively studied in optics, as well as in atomic
physics~\cite{malomedbook}, it also admits stationary points for which the amplitudes of the two oscillators are not equal. To be stationary, such points must satisfy $\theta' = 0$, meaning
\begin{equation}
\tan \theta = \frac{a_1^2-a_2^2}{a_1^2+a_2^2},
\label{p1}
\end{equation}
and be thus associated with angles different from $0$ or $\pi$.
Dividing the amplitude equations of Eq.~(\ref{system}) on both sides by $\cos \theta$, and using Eq.~(\ref{p1}) as well as the trigonometric identity of $\sec^2 \theta =1+\tan^2 \theta$, we obtain two algebraic 
conditions for stationary points characterized by unequal amplitudes:
\begin{align}
3a_2^3 &= \sqrt{a_1^4+a_2^4}\left( \frac{1-r}{\sqrt{2}} a_1+\frac{3}{2\sqrt{2}} s a_1^3 \right) \nonumber \\
3a_1^3 &= \sqrt{a_1^4+a_2^4}\left( \frac{1-r}{\sqrt{2}} a_2+\frac{3}{2\sqrt{2}} s a_2^3 \right).
\label{p2}
\end{align}
The system of Eq.~(\ref{p2}) has to be solved numerically, and the 
resulting values for $a_1$ and $a_2$ can then be substituted into 
Eq.~(\ref{p1}) to obtain the corresponding value of $\theta$. The results 
are shown in Fig.\ref{gensol}(a), where the amplitudes are plotted 
against the gain parameter $r$ for a fixed value of $s=1$. (The qualitative picture does not depend on the precise value of $s$, as long as it is positive.) 
As is evident in the figure, the symmetric solution ($\theta=0$) branches 
off the zero solution at $r=-2$, and the splayed state ($\theta=\pi$) 
branches off at $r=4$, as expected. Importantly, 
another oscillatory pattern emerges after $r=10$, one where the 
two amplitudes, $a_1$ and $a_2$, are unequal. This solution is seen to 
branch off from the splayed (anti-synchronized) state. This is precisely
the (subcritical) pitchfork bifurcation that we referred to previously,
one that decreases by one the number of unstable eigendirections---with
positive real part---of  the splayed state, while producing this novel
symmetry-broken yet unstable branch of solutions. The latter branch
inherits the two eigendirections with a positive real part of
the parent---splayed---branch, in a way consonant with the pitchfork
nature of this symmetry-breaking bifurcation.

\subsection{Nonidentical oscillators}
\label{nonid}

The previous sections assumed that the two oscillators were identical. While electrical components are never truly identical, we would expect those findings to tolerate minor manufacturing variations. If components are deliberately chosen to be different, new phenomena arise.

The analysis for dissimilar components closely follows the one given above, but is more technically and notationally involved. As such, we have placed some of those calculations in Appendix 2. The primary results from that appendix keep track of every single component, but for concreteness we will assume that only the capacitors vary between the oscillators. That is, the two oscillators are as shown in Fig.~\ref{diagram} but with different values $C$ between them, giving rise to two different natural frequencies, $\omega_1$ and $\omega_2$. In that case, we can derive the following governing system:
\begin{align} 
{a_1}^\prime =&  \frac{3}{2}(\omega_2 \cos \theta - \omega_1 \sin \theta) a_2 - \nonumber \\
 &\quad \frac{1}{2}\omega_1(1-r)a_1-\frac{3}{4}\omega_1 s a_1^3 \nonumber \\
{a_2}^\prime =&  \frac{3}{2}(\omega_1\cos \theta +\omega_2 \sin \theta) a_1 - \nonumber \\
 & \quad \frac{1}{2}\omega_2(1-r)a_2-\frac{3}{4}\omega_2 s a_2^3
\nonumber \\
\theta ^\prime =&  \frac{1}{\varepsilon_c}(\omega_1-\omega_2) - \frac{3}{2}\sin \theta \left(\omega_2 \frac{a_2}{a_1}+
\omega_1 \frac{a_1}{a_2}\right)- \nonumber \\
&\frac{3}{2}\cos \theta \left(\omega_1 \frac{a_2}{a_1}-\omega_2 \frac{a_1}{a_2}\right).
\label{general}
\end{align}
Here the average phase velocity $\omega_0$ was absorbed into the non-dimensionalized time variable along with $\varepsilon_c$. The $\omega_{1,2}$ are measured in multiples of $\omega_0$. (Note that the difference in velocities is itself small, comparable to $\varepsilon_c$.) The key difference between this equation and Eq.~(\ref{system}) is the presence of the natural frequencies, which were previously assumed to be identical.

\begin{figure}
\begin{center}
\includegraphics[width=3.3in]{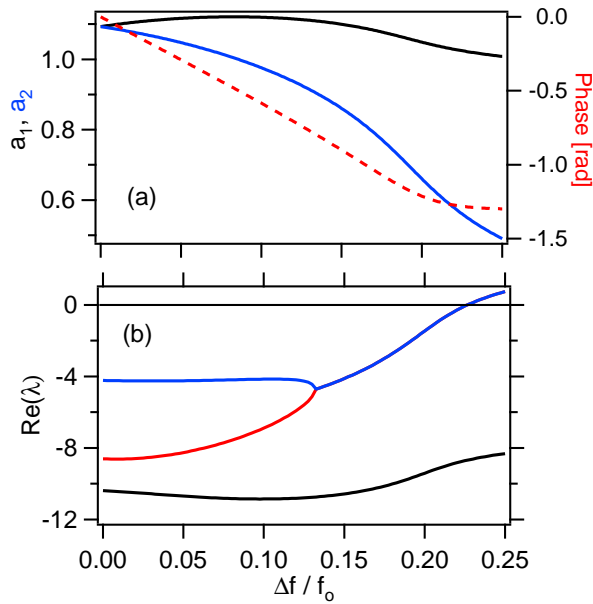}
\caption{(a) The values of $a_1$, $a_2$, and $\theta$ that make the system in Eq.(\ref{general}) stationary. Plotted on the left axis are the amplitudes of the two oscillators versus frequency detuning, and on the right axis the phase difference. $\varepsilon_c=0.05$, $\varepsilon_v=0.24$, $\varepsilon_g=0.33$. (b) The corresponding eigenvalues of the Jacobian for this system. Note that the real part of one eigenvalue crosses zero and becomes positive, indicating a Hopf instability.}
\label{sims2}
\end{center}
\end{figure}

In order to find the fixed points of Eq.~(\ref{general}), we once again 
set the left side to zero and numerically solve for the roots of the system. Figures~\ref{sims2}(a) and (b) illustrate the nature of the fixed points for a specific value of the coupling, $\varepsilon_c = 0.05$.
The steady-state amplitudes and phase depend on frequency mismatch, as demonstrated in Figure~\ref{sims2}(a). As we start detuning one oscillator relative to the other, the amplitudes (solid lines) start to diverge and a phase difference (dashed line) starts to develop. Inserting these fixed-point values into the Jacobian for this system and extracting the eigenvalues yields the picture shown in Fig. \ref{sims2}(b). For low values of $\Delta f / f_o$, three real (negative) eigenvalues are obtained, two of which collide as the de-tuning increases, generating a complex pair. This eigenvalue pair then crosses zero and acquires a positive real part at around $\Delta f/f_o = 0.225$. 
It is at this point that the phase-locked state becomes unstable
via a Hopf bifurcation. While the details of this
bifurcation (and associated periodic orbits) are outside
the scope of our present considerations, monitoring the dynamics
of the phase difference $\theta$ past the relevant
critical point suggests that an increase without bound in
$\theta$, i.e., we observe a phase drift.  
This, in turn, suggests that the relevant Hopf bifurcation
is subcritical (i.e., no stable limit cycle appears to emerge
past the relevant critical point).

Figure~\ref{sims2} has the following physical interpretation. For large detunings the oscillators do not synchronize; for small detunings they synchronize with a phase offset and asymmetric amplitudes. For moderate detunings, the oscillators synchronize, but if perturbed their approach back to steady state would exhibit a ringing effect. For very small detunings, perturbations would exponentially return to the steady-state phase offset and asymmetric amplitudes. 
In the next section, we will focus on the most dramatic and easily observed boundary, that demarcating the synchronized and unsynchronized states. The critical detuning depends upon the coupling strength, $\varepsilon_c$

\section{Comparisons to Experiment}

To evaluate the predictions made during the preceding analysis, we built two Wien-bridge oscillators following the design of Fig.~\ref{diagram}.

The design is nearly identical to those of previous experimental studies (see Refs. \cite{english, tem}), with a key difference in the gain $k\equiv 1 + R_1 / R_2$. Our analysis requires a gain of roughly $k=3$, whereas previous work employed a much larger $k=10$. For smaller gain (roughly less than 5) the diodes merely curb the amplitude of growing harmonic oscillations. For the larger gain, the voltage would grow exponentially were it not for the limiting action of the diodes, and the oscillator is inherently a relaxation oscillator. The high-gain system is difficult to analyze, but the pure exponential growth leads to fairly stiff amplitudes. The amplitude dynamics can then be ignored and the phase oscillator models used to analyze their behavior is well justified. Our analysis relies on the individual oscillators being nearly harmonic, placing our study in a different regime from previous work.

For the preceding theory to apply, $\varepsilon_g$ must be small. In the experiments we chose resistors that gave a value of 0.07, corresponding to a gain marginally above three, i.e., $k=3.07$. The resulting oscillations are very nearly sinusoidal and the two diodes in parallel with $R_1$ only provide very gentle corrections. In this scenario, the assumption of Eq.~(\ref{nonlin}) may be approximately satisfied, as the diodes act to lower the resistance, $R_1$, for larger voltages across it.

Let us begin with a description of the important behaviors. When the two identical oscillators are mutually coupled, they synchronize with identical phases, as predicted. As one of the oscillators is slightly detuned in natural frequency, the synchronization persists, albeit with nonzero phase offset. Beyond a critical frequency mismatch the two oscillators cannot stay locked, and periodic phase slippage ensues, as also reported in Ref.~\cite{english}. In contrast to the behavior in previous studies, the phase slip events are accompanied by noticeable amplitude modulations.

\begin{figure}
\begin{center}
\includegraphics[width=3.4in]{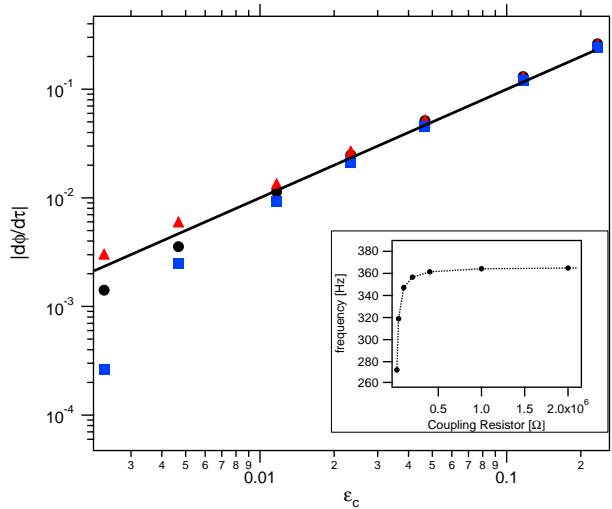}
\caption{The experimental results for the synchronization frequency as a function of coupling strength. In the main panel, we plot the reduced phase velocity as it appears in Eq.(\ref{predic}) against $\varepsilon_c$. The slope is predicted to be 1, as indicated by the solid line. The markers represent experimental data for different capacitor values. Triangles refer to $C= 980 n$F, circles to $C= 96 n$F, and squares to $C= 9.5 n$F. We see that in this representation all data collapses onto the line of slope 1. For very weak coupling, the oscillators are sensitive to imperfect tuning and stray noise. The inset shows the unreduced frequency data as a function of coupling resistance for $C= 96 n$F.}
\label{data}
\end{center}
\end{figure}

To test the theoretical predictions quantitatively, we first measure the dependence of the synchronization frequency on coupling strength. A typical set of measurements is shown in the inset of Fig.~\ref{data}. The coupling resistance varies logarithmically over a range from 20 k$\Omega$ to 4 M$\Omega$. For each value, the frequency of the two synchronized oscillators is measured on an oscilloscope.
Low (high) values of the coupling resistor, $R_c$, correspond to strong (weak) coupling. We see that the measured frequency reaches the uncoupled oscillator frequency asymptotically for large $R_c$.  For lower coupling resistors (stronger coupling), the frequency decreases.

To directly compare this data to Eq.~(\ref{predic}), we recast the $(R_c, f)$ data as $(\varepsilon_c,\dot \phi)$, as shown in the main part of Fig.~\ref{data}. The coupling strengths are obtained from the coupling resistors via $\varepsilon_c = R/R_c$; the phase velocities are obtained from the measured synchronization frequencies via $\dot{\phi} = 2\pi (f-f_0) RC$. Here $R$ denotes the resistance used in the Wien-bridge and held constant at 4.7k$\Omega$. We subtract the uncoupled frequency $f_0$ since $\dot\phi$ characterizes variations from the natural frequency. Figure~\ref{data} includes three different data sets corresponding to three different values of the capacitance, $C$: 9.5 nF (squares), 96 nF (circles), and 980 nF (triangles). Changing the capacitance alters the natural frequencies of the oscillators, letting us sample frequency behavior that differs by two orders of magnitude.
For coupling that is not too weak, the data collapses onto a line with a slope of 1, in excellent agreement with the theory. Since the theory relied on $\pi/4$ as the value for the model-parameter $\alpha$, we have experimentally verified this prediction as well.
\begin{figure}
\begin{center}
\includegraphics[width=3.3in]{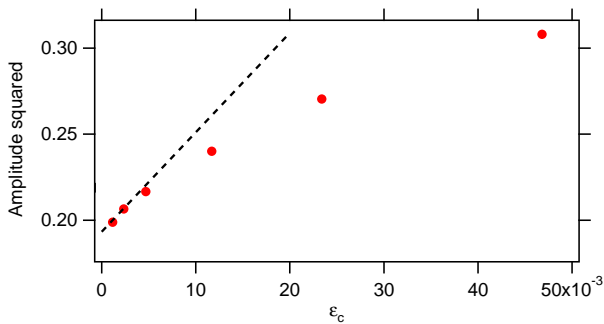}
\caption{The experimental results for the synchronization amplitude squared as a function of coupling strength, $\varepsilon_c$. The dashed line depicts the theoretical prediction from Eq.~(\ref{amp_pred}), whereas the markers represent the measured values. Note that the predicted line has no free parameters. The data for low coupling matches the prediction quite well, but for larger coupling strength the data points veer off from the linear relationship due to the physical properties of the diodes used.}
\label{data2}
\end{center}
\end{figure}

We predicted in Eq.~(\ref{amp_pred}) that the amplitude of the Wien-bridge oscillators would depend upon the coupling strength. The value of $\varepsilon_g$ is known from the ratio of resistances in the voltage-divider part of the circuit, and it evaluates to 0.067. It is difficult to ascertain the value of $\varepsilon_v$ directly from circuit component values, since it appears as a parameter in a heuristic model of the effect of the diodes. However, we can extract it from the measured amplitude when the oscillators are not coupled, for which Eq.~(\ref{amp_pred}) yields $\varepsilon_v=\frac{2\varepsilon_g}{3 a_0^2}$. With that piece of information, all parameters in Eq.~(\ref{amp_pred}) are determined. The resulting prediction for our oscillators is plotted in Fig.~\ref{data2} as the dashed line.

While the amplitude's dependence on the coupling agrees well for small coupling, the range of agreement is somewhat disappointing. The value for $a_v$ is correct for small coupling, which is to be expected since $\varepsilon_v$ was calculated from zero-coupling data. More importantly, the slope of the dependence is also correct for small coupling. To understand the deviations at larger couplings, recall that the effect of the diodes was absorbed into the model as a voltage-dependent resistance in $R_1$, and keeps terms only quadratic in voltage. We know that higher-order terms would also have to be kept for a more realistic diode model and would play a crucial role for higher amplitudes, which is precisely the circumstance where our prediction breaks down.

The squared amplitude results match the theory for very small couplings, while the phase velocity results match the theory for large couplings. Fortunately, the failure of one does not invalidate the other. In particular, $\varepsilon_v$ does not enter into the phase equations, and Eq.~\ref{predic} only depends upon the equality of the amplitudes, not their specific agreement with the predictions. As such, the excellent match between the measured and predicted frequencies are not jeopardized by deviations from the amplitude predictions.

\begin{figure}
\begin{center}
\includegraphics[width=3.2in]{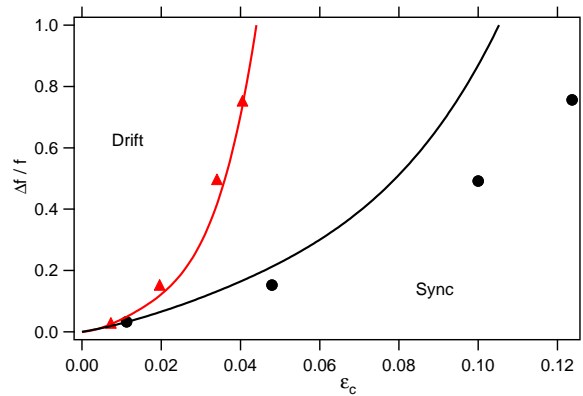}
\caption{The experimental (markers) and analytical (line) phase-boundary between synchronization and phase-drift as a function of coupling strength, $\varepsilon_c$. Triangles (red) show the data for $\varepsilon_g=0.067$, whereas squares (black) correspond to $\varepsilon_g=0.33$. The analytical predictions are obtained by tracking the critical value of $\varepsilon_c$ for which one eigenvalue pair crosses zero and becomes positive.}
\label{data3}
\end{center}
\end{figure}
Finally, we have also experimentally examined the case of two different Wien-bridge oscillators described in section \ref{nonid}. In that section we assumed that the oscillators were identical apart from their capacitors. The simplest testable prediction in this situation is the phase-boundary between the synchronized and drifting states, depicted in Fig.~\ref{data3}. For a gain setting corresponding to $\varepsilon_g=0.067$, the experimental data compares well to the theoretical prediction, as seen by the (red) triangle markers in . Here the four data points were obtained using four different (lower) capacitances in the second oscillator. The choice of $C$ then determines the frequency difference $\omega_2-\omega_1$, and so in order to find the phase boundary, the coupling resistor is varied. As expected, we find that stronger coupling strengths maintain synchronization for higher frequency detunings. The markers closely follow the prediction based on the instability onset of the synchronized state, shown in the figure as the solid (red) line. This instability onset was obtained by recording the zero-crossings of the eigenvalue seen in Fig. \ref{sims2}(b). When the gain setting is increased to $\varepsilon_g=0.33$, the experimental values (black squares) start diverging from the theoretical line for higher coupling strengths. The reason for this divergence at larger $\varepsilon_c$ is most likely the same as already discussed with regard to Fig.~\ref{data2}: larger coupling produces larger oscillation amplitudes, which in turn means that the diode corrections become more prominent.

Interestingly, both the theory and experiment indicate that the phase boundary depends upon the gain. For $\varepsilon_g=0.067$, the minimum difference in natural frequency ($\omega_1-\omega_2$) necessary to break the synchronized state is much larger than for $\varepsilon_g=0.33$, and this trend continues for $\varepsilon_g=0.033$ (not shown). A smaller gain setting (on both oscillators) has the effect of favoring oscillator synchronization.

\section{Conclusions}

We have shown that two coupled low-gain Wien-bridge oscillators can be modeled as a pair of amplitude oscillators with Kuramoto-Sakaguchi-like phase coupling. The equations describing the amplitude and phase dynamics can be derived by applying the method of multiple time scales to the first-principles circuit equations describing the dynamics of voltages and currents in the circuit, and they can be considered a generalization of the phase-oscillator Kuramoto-Sakaguchi model. Simulations indicate that the model's predictions for steady-state behavior agree with the underlying circuit equations. To our knowledge, such a derivation of an amplitude oscillator model from basic physical laws has only been fully accomplished in very limited classes of systems, more specifically two for coupled mechanical oscillators and another in the context of Josephson arrays. A recent study \cite{english} found experimental evidence that electronic self-oscillators were well described by the Kuramoto-Sakaguchi model. While this work does not formally extend to the high-gain regime explored in that study, the analysis does suggest that the experimental findings therein are not unexpected.

The amplitude/phase equations admit a number of interesting stationary states, most of which are unstable. When we assume that the oscillators are identical, the variety of stationary states include equal-amplitude synchronized and anti-synchronized states, as well as and symmetry-broken states of unequal amplitudes. Only the synchronized state featuring constant amplitudes is stable. In this state, the amplitude/phase equations effectively reduce to the Kuramoto-Sakaguchi model with a phase-delay parameter of $\alpha=\nicefrac{\pi}{4}$. When we assume that the oscillators have different frequencies, unsynchronized behavior emerges as a possibility, and we numerically find the boundary between these two states. Focusing on stable steady-state behavior, we obtain a number of measurable predictions for our oscillators.

Some of the salient features of the theory and numerical computation are directly compared to experimental data. Predictions for phase velocity are very good and predictions for the boundary between the synchronized and unsynchronized state are decent. The steady-state amplitude shows the limitations in our method. Overall the amplitude/phase equations shed light on a wide variety of behavior for our oscillators.

Having systematically analyzed the case of the pair paves the way
for numerous future studies. On the one hand, the methodology developed for this ``dimer'' can be generalized to ``oligomer'' settings involving a few
oscillators, e.g., trimers~\cite{roygood}, quadrimers~\cite{zezy} etc.
Another direction is that of extending the present phenomenology to
a full lattice and trying to examine how the synchronization,
phase drifting or symmetry breaking (in the latter setting perhaps 
manifested via localization) features arise. Such studies are currently
under consideration and will be reported in future publications.

\vspace{2mm}

{\it Acknowledgments:}
PGK gratefully acknowledges support from the Alexander von Humboldt Foundation
and from the  US-ARO under grant (W911NF-15-1-0604), as well as from the NSF
under grant DMS-1312856.

\appendix
\section*{Appendix 1: Two-time analysis details}

We write the time derivatives explicitly in terms of the
two time scales, $T_0$ and $T_1$. Defining $\partial_{0}\equiv\partial/\partial T_{0}$
and $\partial_{1}\equiv\partial/\partial T_{1}$, we have 
\begin{align}
\frac{d}{d\tau} & =\partial_{0}+\varepsilon_{t}\,\partial_{1}+\dots\nonumber \\
\frac{d^{2}}{d\tau^{2}} & =\partial_{0}^{2}+2\varepsilon_{t}\,\partial_{0}\partial_{1}+\dots
\end{align}
If we substitute the expansion ansatz of Eq.~(\ref{eq:ansatz}) into
the governing circuit equations, then to zeroth order in small quantities,
we have
\begin{equation}
\partial_{0}^{2}V_{0}+V_{0}=0,
\label{zero}
\end{equation}
and similarly for the other oscillator. This has solutions of the
following form:
\begin{align}
V_{0}(T_{0},T_{1})= & A_{v}(T_{1})\exp(\imath T_{0})+ C.C.\nonumber \\
U_{0}(T_{0},T_{1})= & A_{u}(T_{1})\exp(\imath T_{0})+ C.C.
\label{eq:slow-amplitude-ansatz}
\end{align}
This form is expected in the sense that the fast-time dynamics are
explicitly separated out from the slower amplitude and phase dynamics
of $A_v$ and $A_u$ (both of which are complex). 

Including terms to first order in small quantities, we have from Eq.(\ref{eq:governing-eqs2}),
\begin{align}
0=  \left(\partial_{0}^{2}+2\,\varepsilon_{t}\partial_{0}\partial_{1}\right)V_{in}+\left(\partial_{0}+\varepsilon_{t}\,\partial_{1}\right)V_{in}(\varepsilon_{g}+\varepsilon_{c}\nonumber \\
 +6\,\varepsilon_{v}\,V_{in}^{2})+V_{in}\left(1+\varepsilon_{c}\right)-3U_{in}\varepsilon_{c}-3\left(\partial_{0}+\varepsilon_{t}\,\partial_{1}\right)U_{in}\varepsilon_{c}
\end{align}
We now perform the substitution $V_{in}=V_{0}\left(T_{0},T_{1}\right)+\varepsilon_{1}\,V_{1}\left(T_{0},T_{1}\right)$.
Noting %that %$\partial_{0}^{2}V_{in}+V_{in}=0$ 
Eq.~(\ref{zero}), and keeping terms only
to first order in small quantities, we obtain
\begin{align}
\partial_{0}^{2}V_{1}+V_{1}= & \frac{3\varepsilon_{c}}{\varepsilon_{1}}U_{0}+\frac{3\varepsilon_{c}}{\varepsilon_{1}}\partial_{0}U_{0}-\frac{\varepsilon_{c}}{\varepsilon_{1}}V_{0}\nonumber \\
 & -\frac{2\varepsilon_{t}}{\varepsilon_{1}}\partial_{0}\partial_{1}V_{0}-\frac{\varepsilon_{g}+\varepsilon_{c}+6\varepsilon_{v}\,V_{0}^{2}}{\varepsilon_{1}}\partial_{0}V_{0}.
\end{align}
Note that from Eq.~(\ref{eq:slow-amplitude-ansatz}), we can obtain $V_{0}^{2}\partial_{0}V_{0}=A_v^2 A_v^*\imath e^{\imath T_0}+A_v^2 A_v^*\imath e^{3\imath T_0}+C.C.$,
%inserting what we know about the time dependence of $U_{0}$ and $V_{0}$,
and thus,
\begin{align}
\partial_{0}^{2}V_{1}+V_{1} =\frac{3\varepsilon_{c}}{\varepsilon_{1}}A_{u}\,\mbox{e}^{\imath T_{0}}+\imath\frac{3\varepsilon_{c}}{\varepsilon_{1}}A_{u}\,\mbox{e}^{\imath T_{0}}-\frac{\varepsilon_{c}}{\varepsilon_{1}}A_{v}\,\mbox{e}^{\imath T_{0}} \nonumber \\
-\frac{2\varepsilon_{t}}{\varepsilon_{1}}\imath\partial_{1}A_{v}\,\mbox{e}^{\imath T_{0}}-\frac{\varepsilon_{g}+\varepsilon_{c}+6\varepsilon_{v}\,A_{v}A_{v}^{*}}{\varepsilon_{1}}\imath A_{v}\,\mbox{e}^{\imath T_{0}} \nonumber \\
-\frac{6\varepsilon_v}{\varepsilon_1}A_{v}A_{v}^{*}\imath A_v \mbox{e}^{3\imath T_0}.
\end{align}
We are not interested in actually solving for $V_{1}$. We merely
want to identify the conditions under which the behavior of $V_{1}$
does not diverge, to be consistent with the ansatz of Eq.~(\ref{eq:ansatz}).
All of the terms on the right side, with the exception of the last one, oscillate as $\mbox{e}^{\imath T_{0}}$,
and so represent \emph{resonant} driving. For $V_{1}$ to be stable
these (so-called secular) terms must be zero. That is,
\begin{align}
\frac{3\varepsilon_{c}}{\varepsilon_{1}}A_{u}+\imath\frac{3\varepsilon_{c}}{\varepsilon_{1}}A_{u}-\frac{\varepsilon_{c}}{\varepsilon_{1}}A_{v}-\frac{2\varepsilon_{t}}{\varepsilon_{1}}\imath\partial_{1}A_{v}\nonumber \\
-\frac{\varepsilon_{g}+\varepsilon_{c}+6\varepsilon_{v}\,A_{v}A_{v}^{*}}{\varepsilon_{1}}\imath A_{v}=0.
\label{secular}
\end{align}

To make a connection with the Kuramoto model, we write the slow-varying
complex ``amplitudes'' $A_{v}$ and $A_{u}$ in terms of slowly
varying, purely real amplitude and phase:
\begin{eqnarray}
A_{v}(T_{1}) & =\frac{1}{2}a_{v}(T_{1})e^{\imath \phi_{v}} \nonumber \\
A_{u}(T_{1}) & =\frac{1}{2}a_{u}(T_{1})e^{\imath \phi_{u}}.
\label{ansatz2}
\end{eqnarray}
Substituting into Eq.(\ref{secular}) leads to the following differential equations:
\begin{align}
\partial_{1}a_{v}+\imath a_{v}\,\partial_{1}\phi_{v} =\frac{3\varepsilon_{c}}{2\varepsilon_{t}}\frac{\sqrt{2}}{2}a_{u}\,\mbox{e}^{\imath\left(\phi_{u}-\phi_{v}-\frac{\pi}{4}\right)}\nonumber \\
+\left(\frac{\imath\varepsilon_{c}}{2\varepsilon_{t}}-\frac{2\varepsilon_{g}+2\varepsilon_{c}+3\varepsilon_{v}a_{v}^{2}}{4\varepsilon_{t}}\right)a_{v}.
\end{align}
The real and imaginary parts must hold independently, so
\begin{align}
&\partial_{1}a_{v} =\frac{3\varepsilon_{c}}{2\varepsilon_{t}}\frac{\sqrt{2}}{2}a_{u}\,\cos\left(\phi_{u}-\phi_{v}-\frac{\pi}{4}\right)-\frac{2\varepsilon_{g}+2\varepsilon_{c}+3\varepsilon_{v}a_{v}^{2}}{4\varepsilon_{t}}a_{v}, \nonumber \\
& a_{v}\,\partial_{1}\phi_{v} =\frac{3\varepsilon_{c}}{2\varepsilon_{t}}\frac{\sqrt{2}}{2}a_{u}\,\sin\left(\phi_{u}-\phi_{v}-\frac{\pi}{4}\right)+\frac{\varepsilon_{c}}{2\varepsilon_{t}}a_{v}.
\label{result}
\end{align}
This is the main objective of our study as regards the existence, stability
and dynamics of the different possible states.
If, additionally, the oscillators lock phase and the amplitudes approach a steady-state
value, then by symmetry $a_{v}=a_{u}$ and the dynamics for the phases is given by,
\begin{align}
\partial_{1}\phi_{v}=\frac{1}{2}\frac{\varepsilon_{c}}{\varepsilon_{t}}+\frac{3\sqrt{2}}{2}\frac{\varepsilon_{c}}{\varepsilon_{t}}\sin\left(\phi_{u}-\phi_{v}-\frac{\pi}{4}\right).
\label{SK}
\end{align}
If we multiply by $\varepsilon_t$, and remember the definition of fast and slow times, we arrive at our final (reduced) result:
\begin{align}
\dot{\phi}_{v}=\frac{1}{2}\varepsilon_{c}+\frac{3\sqrt{2}}{2}\varepsilon_{c}\sin\left(\phi_{u}-\phi_{v}-\frac{\pi}{4}\right) \nonumber \\
\dot{\phi}_{u}=\frac{1}{2}\varepsilon_{c}+\frac{3\sqrt{2}}{2}\varepsilon_{c}\sin\left(\phi_{v}-\phi_{u}-\frac{\pi}{4}\right).
\label{SK_final}
\end{align}

\section*{Appendix 2: Unequal oscillator equations}
\label{dissimilar-components}
%The analysis presented in the main body of the paper assumes a number of identical components. 
In this appendix, we relax the assumption of identical components 
and explicitly track each component value.
For the junction at $V_i$ (i=1 or 2, labeling the oscillator) we have $I_{i,out} = I_{ij} + C_{iB} \dot V_i + \frac{V_i}{R_{iB}}$.
Here $I_{i,out}$ is the current flowing into the junction from the Op Amp's output terminal and $I_{ij}$ is the current flowing out of the junction to the output of the other Op Amp, which is at a voltage $V_{j,out}$. The values $C_{iB}$ and $R_{iB}$ are the parallel capacitor and resistor connecting $V_i$ to ground. Using $R_{ij}$ to denote the resistor connecting $V_i$ to $V_{j,out}$, we eliminate the coupling current and obtain
\begin{equation}
I_{i,out} = \frac{V_i - V_{j,out}}{R_{ij}} + C_{iB} \dot V_i + \frac{V_i}{R_{iB}}.
\end{equation}
Denoting with $R_{iT}$ and $C_{iT}$ the resistor and capacitor in series that connect $V_{i,out}$ to $V_i$, Eq.(\ref{eq:almost_main}) can be written as
\begin{align}
0= & R_{iT}C_{iB}\ddot{V}_i + \dot{V}_i \left( 1 + \frac{R_{iT}}{R_{iB}} + \frac{C_{iB}}{C_{iT}}\right) + \frac{V_i}{R_{iB}C_{iT}} \nonumber \\
& + \frac{1}{R_{ij}C_{iT}}\left(V_i-V_{j,out}\right) \nonumber \\
& + \frac{R_{iT}}{R_{ij}}\frac{d}{dt}\left(V_i - V_{j,out}\right)-\dot{V}_{i,out}.
\label{eq:almost_main_dissimilar}
\end{align}
Let us now consider the nonlinear voltage divider branch of a single oscillator. Previously, the resistors were labeled $R_1$ and $R_2$; $R_{01}$ denoted the purely linear response of the nonlinear $R_1$. Here we will call these resistors $R_{iU}$ and $R_{iL}$, and the purely linear term as $R_{0iU}$. The small nonlinearity in the voltage shall be the same, $\varepsilon_v$, so that Eq.(\ref{approx}) becomes
\begin{equation}
V_{i,out} = \left[1 + \frac{R_{0iU}}{R_{iL}} \left(1 - \varepsilon_v V_i^2 \right) \right] V_i.
\end{equation}
Combining this expression with Eq.~(\ref{eq:almost_main_dissimilar}), we arrive at the final voltage equations after
%\begin{align}
%\ddot{V}_{i}= & -V_{i}\left(\frac{1}{R_{ij}R_{iT}C_{iT}C_{iB}}+\frac{1}{R_{iT}R_{iB}C_{iT}C_{iB}}\right)\nonumber \\
% & +\dot{V}_{i}\left(\frac{R_{0iU}}{R_{iL}R_{iT}C_{iB}}-\frac{1}{R_{ij}C_{iB}}-\frac{1}{R_{iB}C_{iB}}\right.\nonumber \\
% & \qquad\quad\left.-\frac{1}{R_{iT}C_{iT}}-3\varepsilon_{v}\frac{R_{0iU}}{R_{iL}R_{iT}C_{iB}}V_{i}^{2}\right)\nonumber \\
% & +V_{j}\frac{1}{R_{ij}R_{iT}C_{iT}C_{iB}}\left(1+\frac{R_{0jU}}{R_{jL}}-\varepsilon_{v}\frac{R_{0jU}}{R_{jL}}V_{j}^{2}\right)\nonumber \\
% & +\dot{V}_{j}\frac{1}{R_{ij}C_{iB}}\left(1+\frac{R_{0jU}}{R_{jL}}-3\varepsilon_{v}\frac{R_{0jU}}{R_{jL}}V_{j}^{2}\right).
%\label{eq:basic-form-rearragned}
%\end{align}
introducing the following helpful constants, 
\begin{align*}
\omega_{i} & \equiv\frac{1}{\sqrt{R_{iT}R_{iB}C_{iT}C_{iB}}},\\
\tau_{i} & \equiv\omega_{i}t,\\
\varepsilon_{ig} & \equiv\frac{R_{0iU}}{R_{iL}}\sqrt{\frac{R_{iB}C_{iT}}{R_{iT}C_{iB}}}-\sqrt{\frac{R_{iT}C_{iT}}{R_{iB}C_{iB}}}-\sqrt{\frac{R_{iB}C_{iB}}{R_{iT}C_{iT}}},\\
\varepsilon_{ijc} & \equiv\frac{R_{iB}}{R_{ij}}.
\end{align*}
Using primes to denote differentiation with respect to $\tau_i$, we then obtain
\begin{align}
V_{i}^{\prime\prime}= & -V_{i}\left(1+\varepsilon_{ijc}\right)\nonumber \\
 & +V_{i}^{\prime}\left(\varepsilon_{ig}-\varepsilon_{ijc}\sqrt{\frac{R_{iT}C_{iT}}{R_{iB}C_{iB}}}\right.\nonumber \\
 & \qquad\quad\left.-3\varepsilon_{v}\frac{R_{0iU}}{R_{iL}}\sqrt{\frac{R_{iB}C_{iT}}{R_{iT}C_{iB}}}V_{i}^{2}\right)\nonumber \\
 & +\varepsilon_{ijc}\left(1+\frac{R_{0jU}}{R_{jL}}\right)\left(V_{j}+V_{j}^{\prime}\sqrt{\frac{R_{iT}C_{iT}}{R_{iB}C_{iB}}}\right).
\label{eq:governing-eqs-dissimilar}
\end{align}
This is the generalized equivalent of Eq.~(\ref{eq:governing-eqs}). %Before proceeding with the method of multiple time scales, we will take a moment to interpret what these definitions and equation mean, and how they differ from the definitions in the main text.
The symbols $\tau_i$, $\varepsilon_{ig}$, and $\varepsilon_{ijc}$ bear a close resemblance to $\tau$, $\varepsilon_g$, and $\varepsilon_c$ in the
identical oscillator case. Previously, the top and bottom resistors and capacitors were assumed equal and identical across oscillators, so that $R_{iB} = R_{iT} \equiv R$ and $C_{iT} = C_{iB} \equiv C$. In that case, the dimensionless timescale $\tau_i$ and deviation in the gain $\varepsilon_{ig}$ of each oscillator reduce to the definitions of $\tau$ and $\varepsilon_g$. The definitions presented here clarify how deviations in individual components effect the oscillator dynamics. In our experiments we change the natural frequency by altering $R_{iT}$, but these definitions suggest that such a modification also effects the oscillator's gain. Modifying an individual circuit component will effect multiple aspects of the oscillator's behavior.

For the method of multiple time scales, we use the ansatz:
\begin{align*}
V_{i}\left(t\right) & \equiv V_{i0}\left(T_{i0},T_{i1}\right)+\varepsilon_{1}V_{i1}\left(T_{i0},T_{i1}\right),\\
T_{i0} & \equiv\tau_{i},\\
T_{i1} & \equiv\varepsilon_{t}\tau_{i}.
\end{align*}
The precise definition of $\varepsilon_{t}$ is not crucial here, it is merely a bookkeeping term for a small quantity. Defining the partial derivatives with respect to these two time scales (suppressing subscripts), we have $\partial_{0} \equiv\frac{\partial}{\partial T_{i0}}, \partial_{1} \equiv\frac{\partial}{\partial T_{i1}}$.
Also note that the derivative of coupled terms will involve $\partial_0 e^{\imath\tau_j}=\frac{dt}{d\tau_{i}}\frac{d}{dt}e^{\imath\omega_{j}t}=\imath\frac{\omega_{j}}{\omega_{i}}e^{\imath\omega_{j}t}.$

To apply the method of multiple time scales, we will assume that timescale defined as $\left(\omega_{j}-\omega_{i}\right)^{-1}$ is comparable to $T_{i1}$. This is equivalent to saying that the spread in natural frequencies is small compared to their average values. As such, $V_{i1}$ then has the form of a harmonic oscillator being driven at a frequency $\omega_i$. Upon using an ansatz analogous to Eq.~(\ref{eq:slow-amplitude-ansatz}) and Eq.~(\ref{ansatz2}), we can arrive at the following amplitude-phase equations:
The explicit time dependence in $\omega_{j}t$ and $\omega_{i}t$ perhaps seems odd, but recall that the method of multiple time scales extracted an oscillatory term of $e^{\imath\omega_{i}t}$. This means that the voltage has the form $V_{i}\left(t\right)=\frac{1}{2}a_{i}e^{\imath\phi_{i}}e^{\imath\omega_{i}t}$. If we define $\theta_{i}\equiv\phi_{i}+\omega_{i}t,\label{eq:theta-definition}$,
then we can write the voltage as $V_{i}\left(t\right)=\frac{1}{2}a_{i}e^{\imath\theta_{i}}$ and rewrite the dynamics in terms of $a_i$ and $\theta_i$: %Also, $\varepsilon_{ijc}$ helped with the bookkeeping of small quantities but now obfuscates the relationship with the original components. Replacing that with the original ratio and splitting out the real and imaginary parts leads to our final phae and amplitude equations:
\begin{widetext}
\begin{align}
\dot{a}_{i}= & \omega_{i}\frac{a_{i}}{2}\left(\varepsilon_{ig}-\frac{R_{iB}}{R_{ij}}\sqrt{\frac{R_{iT}C_{iT}}{R_{iB}C_{iB}}}-3\varepsilon_{v}\frac{R_{0iU}}{R_{iL}}\sqrt{\frac{R_{iB}C_{iT}}{R_{iT}C_{iB}}}\frac{a_{i}^{2}}{4}\right)\nonumber \\
 & +\frac{R_{iB}}{R_{ij}}\left(1+\frac{R_{0jU}}{R_{jL}}\right)\frac{a_{j}}{2}\left[\omega_{i}\sin\left(\theta_{j}-\theta_{i}\right)+\omega_{j}\sqrt{\frac{R_{iT}C_{iT}}{R_{iB}C_{iB}}}\cos\left(\theta_{j}-\theta_{i}\right)\right],\label{eq:amplitude-dynamics}\\
a_{i}\dot{\theta}_{i}= & \omega_{i}a_{i}+\frac{R_{iB}}{R_{ij}}\frac{a_{i}}{2}\omega_{i}
  +\frac{R_{iB}}{R_{ij}}\left(1+\frac{R_{0jU}}{R_{jL}}\right)\frac{a_{j}}{2}\left[\omega_{j}\sqrt{\frac{R_{iT}C_{iT}}{R_{iB}C_{iB}}}\sin\left(\theta_{j}-\theta_{i}\right)-\omega_{i}\cos\left(\theta_{j}-\theta_{i}\right)\right].\label{eq:phase-dynamics}
\end{align}
\end{widetext}
In order for this to reduce to Eq.~(\ref{general}), we need to impose the following experimental restrictions. All of the resistors in the RC branch must be identical across both oscillators, so $R_{iT} = R_{iB} \equiv R$. The resistors in the voltage divider must not depend upon oscillator, in which case $\varepsilon (R_{ojU} / R_{jL}) = 2\varepsilon$ to first order in small quantities. The coupling resistors must be the same, $R_{ij} \equiv R_c$. The top and bottom capacitors must be the same within a given oscillator, but we obtain different natural frequencies by letting them differ between oscillators: $C_{1T} = C_{1B} \equiv C_1 \ne C_{2T} = C_{2B} \equiv C_2$.
\end{document}